\documentclass[onecolumn,preprintnumbers,amsmath,amssymb]{revtex4}

\usepackage[dvips]{graphicx}
\usepackage{graphicx}
\usepackage{dcolumn}
\usepackage{bm}
\usepackage{amsmath}
\usepackage{amsfonts}
\usepackage{amssymb}
\usepackage{color}
\usepackage{pstcol}
\usepackage[all]{xy}
\def\be{\begin{equation}}
\def\ee{\end{equation}}

\begin{document}


\title{Quantal Definition of the Weak Equivalence Principle}

\author{Abel Camacho}
\email{acq@xanum.uam.mx} \affiliation{Departamento de F\'{\i}sica,
Universidad Aut\'onoma Metropolitana--Iztapalapa\\
Apartado Postal 55--534, C.P. 09340, M\'exico, D.F., M\'exico.}

\author{ Arturo Camacho--Guardian}
\email{40700163@escolar.unam.mx}
\affiliation{ Facultad de Ciencias\\
Universidad Nacional Aut\'onoma de M\'exico.}


\date{\today}

\begin{abstract}
The present work analyzes the meaning of the Weak Equivalence
Principle in the context of quantum mechanics. A quantal
definition for this principle is introduced. This definition does
not require the concept of trajectory and relies upon the phase
shift induced by a gravitational field in the context of a quantum
interference experiment of two coherent beams of particles. In
other words, it resorts to wave properties of the system and not
to classical concepts as the idea of trajectory.
\end{abstract}


\maketitle
\section{Introduction}
The modern idea of gravitation, at least in its classical
description, is based upon the postulates of metric theories
\cite{Will1}, which embody the so--called Weak Equivalence
Principle (WEP). This principle postulates: {\it If an uncharged
test body is placed at an initial event of space--time and given
an initial velocity, then its subsequent trajectory will be
independent of its internal structure and composition}
\cite{Will1}.

This principle has its experimental foundation in the universality
of free fall, an experiment first performed by Galileo
\cite{Gaga1}, and it entails a relation between two different
concepts of mass, i.e., inertial mass ($m_{(i)}$) and passive
gravitational mass ($m_{(p)}$). Indeed, the classical equation of
motion for a point--like particle immersed in a homogeneous
gravitational field (along the $z$--axis) reads

\begin{eqnarray}
m_{(i)}\frac{d^2 \vec{r}}{dt^2} = -m_{(p)}g\hat{z}\label{WEPCla1}.
\end{eqnarray}

WEP states that the solution to this equation depends upon initial
conditions, but not on the internal structure of the considered
particle. In other words, it is based upon the fact that

\begin{eqnarray}
\frac{m_{(i)}}{m_{(p)}} =\alpha\label{WEPCla2},
\end{eqnarray}

is a universal constant, i.e., the same for all particles.

At this point two remarks have to be done. Firstly, WEP requires
as a necessary condition a particle--independent value of
$\alpha$. In other words, the particular value of $\alpha$ is
completely irrelevant. This may be understood if we introduce
(\ref{WEPCla2}) into the motion equation

\begin{eqnarray}
\frac{d^2 \vec{r}}{dt^2} = \tilde{g}\hat{z}\label{WEPCla3}.
\end{eqnarray}

Here $\tilde{g}=g/\alpha$. A value of $\alpha\not =1$ would entail
a redefinition of the acceleration of gravity. In an experiment we
cannot detect $g$ or $\alpha$, only the {\it effective}
acceleration of gravity, i.e., $\tilde{g}=g/\alpha$. This proves
that the value of $\alpha$ (assuming the validity of WEP) is
irrelevant, it cannot be measured. Secondly, the geometrization of
gravity requires WEP. In other words, if $\alpha$ is
particle--dependent, then WEP breaks down and therefore the
geometrization of gravity would be a dubious issue, at least with
a four--dimensional manifold. Finally, let us add that any
experimental test of the validity, or invalidity, of WEP requires
the use of more than one particle. Indeed, WEP is based upon the
universal character of $\alpha$ and one particle cannot provide us
this kind of information.

 The mathematical version of WEP is done stating: {\it The motion
 of an uncharged particle is provided by the geodesics of the corresponding
manifold} \cite{Will1}. Notice that behind this phrase lies the
concept of trajectory. The experimental verification of WEP in the
classical level has been done in different approaches and devices
\cite{Haugan1} but the quantum--mechanical version demands a
careful analysis. This last remark acquires relevance since in
quantum theory the concept of trajectory is not a necessary one,
i.e., the theory can be formulated without it \cite{Sakurai1}. We
may cast this last idea in the following manner: Is it
conceptually correct to formulate WEP in the validity region of
quantum theory in terms of a concept which happens to be not a
fundamental one in quantum theory? This last comment leads us to
the main question in the present work. Indeed, since WEP is
formulated in terms of trajectories and quantum mechanics does not
contain this notion as a central one, how could we formulate WEP
in terms of ideas fundamental in quantum theory? The present work
puts forward a definition in this sense. This will be done
resorting to the idea of interference pattern, which is a
wave--like property.

\bigskip
\bigskip

\section{Gravity--Induced Phase Shifts and WEP}
\bigskip
\bigskip

\begin{figure}[h!]
{\centerline{ \mbox{
\includegraphics[width=90mm,angle=0,keepaspectratio]{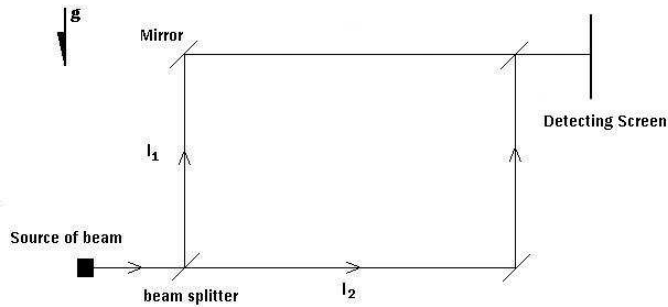}}}}
\caption{ Experimental device. } \label{f-1}
\end{figure}

\subsection{Phase--Shifts and Gravity}
\bigskip

The first experiments (usually denoted by COW) showing how gravity
appears in the realm of quantum theory can be considered to begin
in 1974 \cite{Colella1, Colella2}. The phenomenon dealt with in
this series of experiments can be denoted a gravity--induced
quantum interference. They embody an interference pattern of
thermal neutrons induced by gravity. A brief description of this
experiment is the following one \cite{Colella1, Colella2}.  The
idea is to use an almost monoenergetic beam of thermal neutrons.
This last statement means a kinetic energy of about 20 MeV, which
is tantamount to a speed of $2000ms^{-1}$. The primary beam is
split into two parts, such that each one of the new beams travels
along different paths. These paths define a parallelogram of sides
$l_1$ and $l_2$ (see Fig. 1). If these two paths lie in a
horizontal plane, then, there is vanishing relative phase shift
between the two beams. If we now rotate the plane, say an angle
$\theta$ around any of its arms, then a non--vanishing relative
phase shift appears, due to the fact that the paths, these
secondary beams follow, are located at heights associated with
different gravitational potential. It can be shown that the phase
difference between the two secondary beams, $\Delta\phi$, has the
following form \cite{Colella1, Colella2, Colella3, Werner1,
Colella4}

\begin{eqnarray}
\Delta\phi =
\frac{m^2gl_1l_2\lambda\sin(\theta)}{\hbar^2}\label{Atomic3}.
\end{eqnarray}

In this last expression $m$ and $\lambda$ denote the mass of the
neutrons and the de Broglie wavelength of the neutrons,
respectively. Notice that in all these experiments no distinction
is introduced between inertial and passive gravitational mass
\cite{Colella3, Colella4}.  The dependence of (\ref{Atomic3}) on
the mass of the involved particles has originated a hot debate
about the possible presence of a non--geometric element of gravity
at the quantum realm, and in consequence the claim of a violation
of WEP has been put forward \cite{Ahluwalia1, Ahluwalia2}. This
ambiguity of COW in relation with WEP appears even in some texts
books on quantum mechanics, see, for instance, the last paragraph
and footnote on page 129 in \cite{Sakurai1}.

Of course, we may find in the literature the opposite position.
For instance, already in the abstract of \cite{Mannheim1} we may
find the following phrase: {\it We show that the COW experiment
respects the equivalence principle}. This work \cite{Mannheim1} is
an interesting one and sheds light upon some similarities between
the classical and quantal versions of the experiment.

Other results within the context of neutron interferometry address
different issues related to this experiment. For instance, the
influence of spacetime torsion upon these experiments has already
been analyzed \cite{Ausdretsch1}, as well as the general
relativistic description of them \cite{Ryder1}. Nevertheless,
these works do not tackle the relation between the mass dependency
of the phase shift and WEP. As a matter of fact, a fleeting glance
at these works allows  us to confirm that none of them establishes
a distinction between inertial and passive gravitational mass. In
other words, they take by granted WEP.

Concerning this debate around the validity of WEP in the context
of quantum theory we will take no sides. The appearance of mass in
expression (\ref{Atomic3}) seems to the authors neither enough to
state that WEP is violate, nor fulfilled. The situation calls for
some new solution that could lead to the disentanglement of the
problem.

Our contribution lies along this direction. One possibility has
already been considered, at least theoretically, i.e., free fall
of quantum particles, with and without classical analogue
\cite{Onofrio1}. Nevertheless, as pointed out  the use of this
idea seems to imply a redefinition of WEP for quantum theory since
no quantum object can reproduce the classical concept of
deterministic trajectory and the macroscopic version of WEP is
formulated in terms of trajectories \cite{Onofrio1}.

The present proposal is a slight generalization of COW, in order
to consider the possibility of testing directly a quantal version
for WEP. This definition does not require the concept of
trajectory and relies upon the phase shift induced by a
gravitational field in the context of a quantum interference
experiment of two coherent beams of particles, i.e., it resorts to
wavelike properties. Additionally, an experimental proposal
designed to test it is also here put forward.
\bigskip
\bigskip

\subsection{Quantum WEP}
\bigskip

We will address the issue from its most fundamental aspect.
Firstly, as mentioned in the introduction WEP is based upon the
fact that $m_{(i)}/m_{(p)}$ is particle independent. This
condition implies the universality of free fall. Secondly, quantum
mechanics does not require the concept of trajectories. In other
words, a quantal version of WEP based upon the idea of trajectory
seems conceptually a mistake. Hence, the problem we face is to
define in a consistent way the meaning of WEP.

A possibility is to define WEP operationally as follows. Let us
re-write the phase shift introducing from square one the
distinction between inertial and passive gravitational masses.
Then we obtain

\begin{eqnarray}
\Delta\phi =
\frac{m_{(i)}m_{(p)}gl_1l_2\lambda\sin(\theta)}{\hbar^2}\label{Atomic4}.
\end{eqnarray}

Introducing $\frac{m_{(i)}}{m_{(p)}} =\alpha$ we have

\begin{eqnarray}
\alpha =
\frac{m_{(i)}^2gl_1l_2\lambda\sin(\theta)}{\hbar^2\Delta\phi}\label{Atomic5}.
\end{eqnarray}

This expression allows us to put forward a quantal WEP definition.
Indeed, a fleeting glance at (\ref{Atomic5}) shows us that now we
may distinguish between different particles, in the sense that
$\alpha$ could be particle--dependent. The usual expression, see
(\ref{Atomic3}), assumes that $\alpha =1$ for all particles, and
in consequence assumes from the very beginning the validity of
WEP.
\bigskip

{\bf Definition}: Let us consider a COW device, i.e., the geometry
of it is already given. This means $l_1$, $l_2$ the lengths of the
arms, and the tilt angle $\theta$ are fixed  parameters, and also
$g$. WEP in the quantum realm means that

\begin{eqnarray}
\alpha =
\frac{m_{(i)}^2gl_1l_2\lambda\sin(\theta)}{\hbar^2\Delta\phi}\label{Atomic6},
\end{eqnarray}

is a universal a parameter, i.e., independent of the mass of the
employed particles. Before proceeding let us mention that this
definition is motivated by two factors. Firstly, as mentioned
before the core behind the classical WEP is encoded in a universal
value for $\alpha$, and in this sense our definition embodies the
most substantial factor involved in the macroscopic version of
this principle. In other words, our definition encompasses the
main ingredient behind the macroscopic version of WEP. Secondly,
the idea is to have a feasible experimental verification of it.
Since our principle resorts to COW we need a slight generalization
of it to have a possibility to falsify our definition.

\subsection{Experimental Proposals}
\bigskip

At our disposal, as experimental parameters, we have the arms
length of the interferometer, the wavelength of the particle, and
the tilt angle. This remark entails us to introduce three
different variations of the same kind of experiment.

\subsubsection{First Possibility: Wavelength}
\bigskip

The first series of experiments modifies only the wavelength of
the particle.  With our arm lengths $l_1$ and $l_2$ and a certain
non--vanishing tilt angle $\theta$ we tune the velocity of our
first type of particles (whose inertial mass will be denoted by
$m^{(1)}_{(i)}$). Hence, since momentum and wavelength satisfy de
Broglie's relation, i.e., $\lambda =\hbar/p$, (\ref{Atomic5})
becomes

\begin{eqnarray}
\alpha^{(1)} = \frac{m^{(1)}_{(i)}gl_1l_2\sin(\theta)}{\hbar
v\Delta\phi}\label{Atomic7}.
\end{eqnarray}

Here $\alpha^{(1)}$ takes into account the possible
particle--dependence of this parameter. Suppose now that we
perform our experiment in such a way that our first case for the
velocity, $v_{(1)}$, provides a constructive interference pattern,
hence $\Delta\phi =2\pi n$, where $n$ is an integer. Our second
experiment employs a velocity, $v_{(2)}$, such that it provides
the nearest destructive interference pattern to the first
experiment (there are two velocities, one above $v_{(1)}$ and one
below it fulfilling this condition). Then we have the following
two expressions

\begin{eqnarray}
2\alpha^{(1)}\pi n= \frac{m^{(1)}_{(i)}gl_1l_2\sin(\theta)}{\hbar
v_{(1)}}\label{Atomic8},
\end{eqnarray}

\begin{eqnarray}
\alpha^{(1)}\Bigl[2\pi n \pm\pi\Bigr] =
\frac{m^{(1)}_{(i)}gl_1l_2\sin(\theta)}{\hbar
v_{(2)}}\label{Atomic9}.
\end{eqnarray}

With them we obtain

\begin{eqnarray}
\alpha^{(1)}\pi =
\frac{m^{(1)}_{(i)}gl_1l_2\sin(\theta)}{\hbar}\sqrt{\Bigl[\frac{1}{v_{(2)}}
- \frac{1}{v_{(1)}}\Bigr]^2}\label{Atomic10}.
\end{eqnarray}

We now carry out the same experiment with a different kind of
particles ($m^{(2)}_{(i)}$) and obtain

\begin{eqnarray}
\alpha^{(2)}\pi =
\frac{m^{(2)}_{(i)}gl_1l_2\sin(\theta)}{\hbar}\sqrt{\Bigl[\frac{1}{w_{(2)}}
- \frac{1}{w_{(1)}}\Bigr]^2}\label{Atomic11}.
\end{eqnarray}

Here $w_{(2)}$ and $w_{(1)}$ are the corresponding velocities,
which do not need to be equal to $v_{(2)}$ and $v_{(1)}$. Notice
that in this way we may deduce the value of $\alpha^{(n)}$,
$n=1,2$. We may now put forward an experimental proposal for
testing the just introduced quantal WEP definition. Let us assume
that we perform several times, resorting to different kind of
particles, this experiment. According to our definition $\alpha$
has to be particle--independent, hence, if WEP is to be fulfilled,
then for different particles the same value for $\alpha$ shall
emerge. More than one value would imply the breakdown of WEP.

\subsubsection{Second Possibility: Arm Length}
\bigskip

If the tuning up of the velocity becomes a nuisance, then the
possibility of having several samples of the silicon crystal
interferometer could be exploited, and in this way we obtain an
expression tantamount to $(\ref{Atomic10})$. One of the arms, say
$l_1$ takes to different values. From now on we write for these
two values $l^{(i)}_1$, where, $i=1,2$. This means that in this
situation we have (here the velocity $v$ is the same for both
experiments)

\begin{eqnarray}
\alpha^{(1)}\pi = \frac{m^{(1)}_{(i)}gl_2\sin(\theta)}{v\hbar}\mid
l^{(1)}_1 - l^{(2)}_1\mid\label{Atomic12}.
\end{eqnarray}

Clearly, in this manner we may compare the value of $\alpha$ for
different type of particles.

\subsubsection{Third Possibility: Acceleration of Gravity}
\bigskip

The third possibility could be the easiest to handle. It
encompasses different values for the acceleration of gravity. This
does not mean, necessarily, the use of satellites, though this is
also an additional possibility. A fleeting glance to
$(\ref{Atomic3})$ shows us that in COW we have an {\it effective
acceleration of gravity} given by

\begin{eqnarray}
\tilde{g} = g\sin(\theta)\label{Gravity}.
\end{eqnarray}

Then we may {\it change the effective acceleration of gravity}
with $\theta$. This means that now we have two different values
for our effective acceleration $\tilde{g}^{(i)}$, $i=1, 2$ such
that

\begin{eqnarray}
\alpha^{(1)}\pi = \frac{m^{(1)}_{(i)}l_1l_2}{v\hbar}\mid
\tilde{g}^{(1)} - \tilde{g}^{(2)}\mid\label{Atomic13}.
\end{eqnarray}

Once again we may use our derived expression with different kind
of particles and test our quantal WEP definition.

We may wonder what information could be elicited from
\cite{Colella2} in connection with the present model. Figure 2 on
page 1473 allows us to estimate (using expression (2) on 1472) in
a rough way the value of $\alpha$ for neutrons. Indeed, consider
the first maximum and minimum to the right of $\phi=0$. The angle,
in degrees, of the crest is, approximately, 2, while the valley is
around 5 degrees. The experimental parameters are all given except
the value of the acceleration of gravity, i.e., $g$. Taking a
value of $g=9.83m/s^2$ we obtain $\alpha=0.99951$ for the case
depicted in \cite{Colella2}. Comparing against the result obtained
in \cite{Koester1} ($\alpha\in [0.99991, 1.00041]$) we conclude
that our rough procedure provides a reasonable value for $\alpha$.

\bigskip
\bigskip

\section{Conclusions}
\bigskip
\bigskip

A simple definition for a quantal WEP version has been put
forward, the one does not require the concept of trajectory and
that also includes for macroscopic bodies the main element behind
this principle. Our proposal is connected with the phase shift
appearing in COW and measures a possible particle dependency in
the relation inertial mass--passive gravitational mass. Three
different experiments have been also mentioned.

Finally, let us address an additional issue which has to be
contemplated in the context of the feasibility of the present
proposal. As underlined before, different kinds of particles are
required in order to test WEP, i.e., neutron interferometry does
not suffice. Therefore, we must mention some other systems that
could be employed in a interference experiment of this sort. The
first atomic interferometry experiment is already an old
experiment \cite{Carnal1}, though in this case it is a Young
type--like device, i.e., it measures space coherence, and not time
coherence as in a COW experiment, which can be catalogued as a
Mach--Zehnder situation \cite{Chartier1}. Additional results may
already be found in the extant literature \cite{Keith1, Kasevich1,
Borde1, Borde2}. These last works prove that atomic and even
molecular interferometry are already a real possibility. These
devices introduce different sort of particles, a factor which is
the key element in our work. In other words, we may contemplate
the possibility of resorting to atomic interferometry in the
detection of gravity--induced phase shifts. Of course, this
requires additional work, and at this point it defines an
experimental problem.

Finally, modern elementary particle theories regard, for instance,
the neutron as a system embodying more than one elementary
particle called quarks \cite{Greiner}. In our experiment we do not
take into account this structure. The reason for it is quite
simple, namely, the quark confinement hypothesis, i.e., free
quarks are never seen in isolation. Indeed, the quark confinement
hypothesis, which can be understood as an implication of the
asymptotic freedom, tells us that the force between two quarks
increases at lower energies as they are separated more and more
from one another. In other words, it seems that a possible
detection of the quarks, defining a neutron, by COW could imply a
violation of the quark confinement hypothesis.

\begin{acknowledgments}
 This research was  supported by CONACYT Grant 47000--F and by the Mexico--Germany collaboration
 grants CONACyT--DFG J110.491/2006 and J110.492/2006. The author would
like to thank A. A. Cuevas--Sosa for useful discussions and
literature hints.
\end{acknowledgments}

{}
\end{document}